\documentclass[aps,prl,10pt,twocolumn,showpacs,titlepage,superscriptaddress]{revtex4-1}
\usepackage{graphicx}
\usepackage{color}
\usepackage[colorlinks, linkcolor={blue},citecolor={blue}, urlcolor={blue}]{hyperref}

\begin{document}
\date{\today}

\title{Microwave Experiments on Quantum Search and Directed Transport in Artificial Graphene}
\author{Julian B\"{o}hm}
\author{Matthieu Bellec}
\author{Fabrice Mortessagne}
\author{Ulrich Kuhl}
\email{ulrich.kuhl@unice.fr}
\affiliation{Laboratoire de Physique de la Mati\`ere Condens\'ee, UMR 7336, Universit\'e Nice Sophia Antipolis, CNRS, Parc Valrose, 06100 Nice, France}
\author{Sonja Barkhofen}
\affiliation{AG Quantenchaos, Fachbereich Physik der Philipps-Universit\"at Marburg, D-35032 Marburg, Germany}
\affiliation{Applied Physics, University of Paderborn, Warburger Strasse 100, 33098 Paderborn, Germany}
\author{Stefan Gehler}
\affiliation{AG Quantenchaos, Fachbereich Physik der Philipps-Universit\"at Marburg, D-35032 Marburg, Germany}
\affiliation{Department of Energy Management and Power System Operation, University of Kassel, D-34121 Kassel, Germany}
\author{Hans-J\"urgen St\"ockmann}
\affiliation{AG Quantenchaos, Fachbereich Physik der Philipps-Universit\"at Marburg, D-35032 Marburg, Germany}
\author{Iain Foulger}
\author{Sven Gnutzmann}
\author{Gregor Tanner}
\affiliation{School of Mathematical Sciences, University of Nottingham, University Park, Nottingham NG7 2RD, United Kingdom}

\begin{abstract}
A series of quantum search algorithms have been proposed recently providing an algebraic speedup compared to classical search algorithms from $N$ to $\sqrt{N}$, where $N$ is the number of items in the search space.
In particular, devising searches on regular lattices has become popular in extending Grover's original algorithm to spatial searching.
Working in a tight-binding setup, it could be demonstrated, theoretically, that a search is possible in the physically relevant dimensions 2 and 3 if the lattice spectrum possesses Dirac points.
We present here a proof of principle experiment implementing wave search algorithms and directed wave transport in a graphene lattice arrangement.
The idea is based on bringing localized search states into resonance with an extended lattice state in an energy region of low spectral density---namely, at or near the Dirac point.
The experiment is implemented using classical waves in a microwave setup containing weakly coupled dielectric resonators placed in a honeycomb arrangement, i.e., artificial graphene.
Furthermore, we investigate the scaling behavior experimentally using linear chains.
\end{abstract}


\maketitle


\emph{Introduction.}---Currently, one of the most fruitful branches of quantum information is the field of quantum search algorithms.
It started with Grover's work \cite{gro97} describing a search algorithm for unstructured databases, which has been
implemented experimentally in NMR \cite{chu98,jon98} and in optical experiments \cite{wal05}.
More recently, spatial quantum search algorithms have been proposed based on the quantum walk
mechanism \cite{SKW03,Por13}. All of these algorithms can achieve up to quadratic speedup compared to the
corresponding classical search. For quantum searches on generic $d$-dimensional lattices certain restrictions have
been observed, however, depending on whether the underlying quantum walk is discrete \cite{ADZ93} or continuous
\cite{FG98}. While effective search algorithms for discrete walks on square lattices have been reported for
$d \ge 2$ \cite{AKR05,ADMP10}, continuous-time quantum search algorithms on the same lattice show speedup
compared to the classical search only for $d\ge 4$ \cite{CG04}. Experimental implementations of discrete quantum
walks need time stepping mechanisms such as laser pulses \cite{Kar09,Sch09,Xue09,Zae10,Schr10,sch12}.
By switching to a continuous-time evolution based, for example, on tight-binding coupling between sites, one can
avoid time discretization in an experiment. It has been shown in Ref.~\cite{FGT14} that continuous-time quantum search
in 2D is indeed possible when performed near the Dirac point in graphene or, more generally, for lattices with a cone
structure in the dispersion relation \cite{CG14}, i.e., a linear growth of the density of states (DOS). This effect adds a
new dimension to the material properties of graphene \cite{CN09,WA11} with potential applications in sensing and
detection as well as directed charge carrier transport. This may provide new ways of channeling intensity and information
 across lattices and between distinct sites, such as for single-molecule sensing, as described in Refs.~\cite{Sch07,Weh08}.

In this Letter, we present the first {\em proof-of-principle} experiment for a continuous 2D search in a tight-binding
setup based on a microwave experiment using artificial graphene as discussed in Refs.~\cite{kuh10a,bel13a,bel13b}.
As was already noted by Grover and co-workers, ``quantum'' searching is often a pure wave phenomenon
based on interference alone and can thus also be implemented using the single particle Schr\"{o}dinger equation \cite{gro01} or classical waves such as coupled harmonic oscillators \cite{gro02,rom07} or wave optics \cite{bha02}.
A wave search amongst $N$ sites will take place in a full $N$-dimensional state space, while a full quantum search can be
implemented with only $\log N$ particles. The $\sqrt{N}$ speed-up is independent of this resource
compression issue.

In the following, we recapitulate briefly the theory of quantum searching on graphene and describe the experimental setup.
We then demonstrate both searching and directed transport in graphenelike lattice structures. The $\sqrt{N}$ scaling behavior with a number of sites will be demonstrated experimentally using linear chains.


\emph{Theoretical background.}---All quantum search algorithms starting from Grover's search on an unstructured database are based on the same principle:
the system is set up by bringing an extended state of the (unperturbed) system into resonance with a localized state originating from a perturbation, thereby forming an avoided crossing in the spectrum of resonances.
One then uses the two-level dynamics at the avoided crossing in order to rotate the system from the extended state into the localized state, thus ``finding'' the position of the perturbation \cite{hei09b,HT09}.
The subtleties in setting up such a search lie in
(i) choosing an unperturbed system with eigenstates extending uniformly across all sites,
(ii) finding a suitable perturbation which carries a localized state, and
(iii)  working in an energy range with a low density of states, making it possible to isolate the two-level crossing from the rest of the spectrum.
Note that the starting state can also be a localized state which couples into the search state via an extended state as in the quantum state transfer setup described in Ref.~\cite{HT09} or the continuous Grover search algorithm presented in Refs.~\cite{rom06,rom07,rom09}.

In the following, we will focus on continuous-time walks on regular, finite lattices in a tight-binding setup.
The entries in the database are represented by the $N$ lattice sites and associate orthonormal states  $\{ |i\rangle\}_{i=1}^N$ spanning an $N$-dimensional Hilbert (search) space.
The system is described by a tight-binding Hamiltonian $H_0$ with associate eigenstates $\{|e\rangle\}_{e=1}^N$.
The extended eigenstates most suitable for a search are those corresponding to reciprocal lattice vectors at $k=0$ or near the band edge.
We introduce a perturbation of the form $H= H_0 +W$,
with $W$  supporting a localized state at the ``marked site'' being in resonance with a uniform extended state.
The interaction at the avoided crossing is controlled by the overlap integral $|\langle i | e\rangle|^2 \approx N^{-1}$, which leads to an energy gap of order $1/\sqrt{N}$.
The search is started by preparing the system in an extended state $|e\rangle$.
The state then evolves according to $|\psi(t) \rangle=e^{-iHt/\hbar} |e\rangle$, describing a rotation
taking place predominantly in the two-dimensional subspace spanned by $| i\rangle, |e\rangle$.
Performing a measurement after  the system has evolved into $| i\rangle$ at a time $t_c = {\cal O}(\sqrt{N})$ completes the search.

For generic lattices, the number of states with energy less than $E$ in $d$ dimensions scales typically as $N E^{d/2}$.
One thus expects the first state above the ground state to be at an energy $E_1= O(N^{- 2/d})$.
An energy separation between the states at the avoided crossing [being at a distance $O(N^{-1/2})$] and the rest of the spectrum is thus only possible for $d\ge 4$ in the large $N$ limit.
Running the search at a Dirac point overcomes this problem;
here the energy scaling is reduced to  $E_1 = O(N^{-1/d})$ and searching becomes possible for $d \ge 2$;
the search time is then proportional to $\sqrt{N\ln N}$, as shown in the theoretical studies in Refs.~\cite{FGT14, CG14}.
In contrast to the treatment in Ref.~\cite{FGT14}, the perturbation $W$ introduced in this Letter---is in terms of coupling an additional site to the lattice state $| i\rangle$ with the perturber's on-site energy---tuned to an extended state close to the Dirac point.

\begin{figure}
  \includegraphics[width=\linewidth]{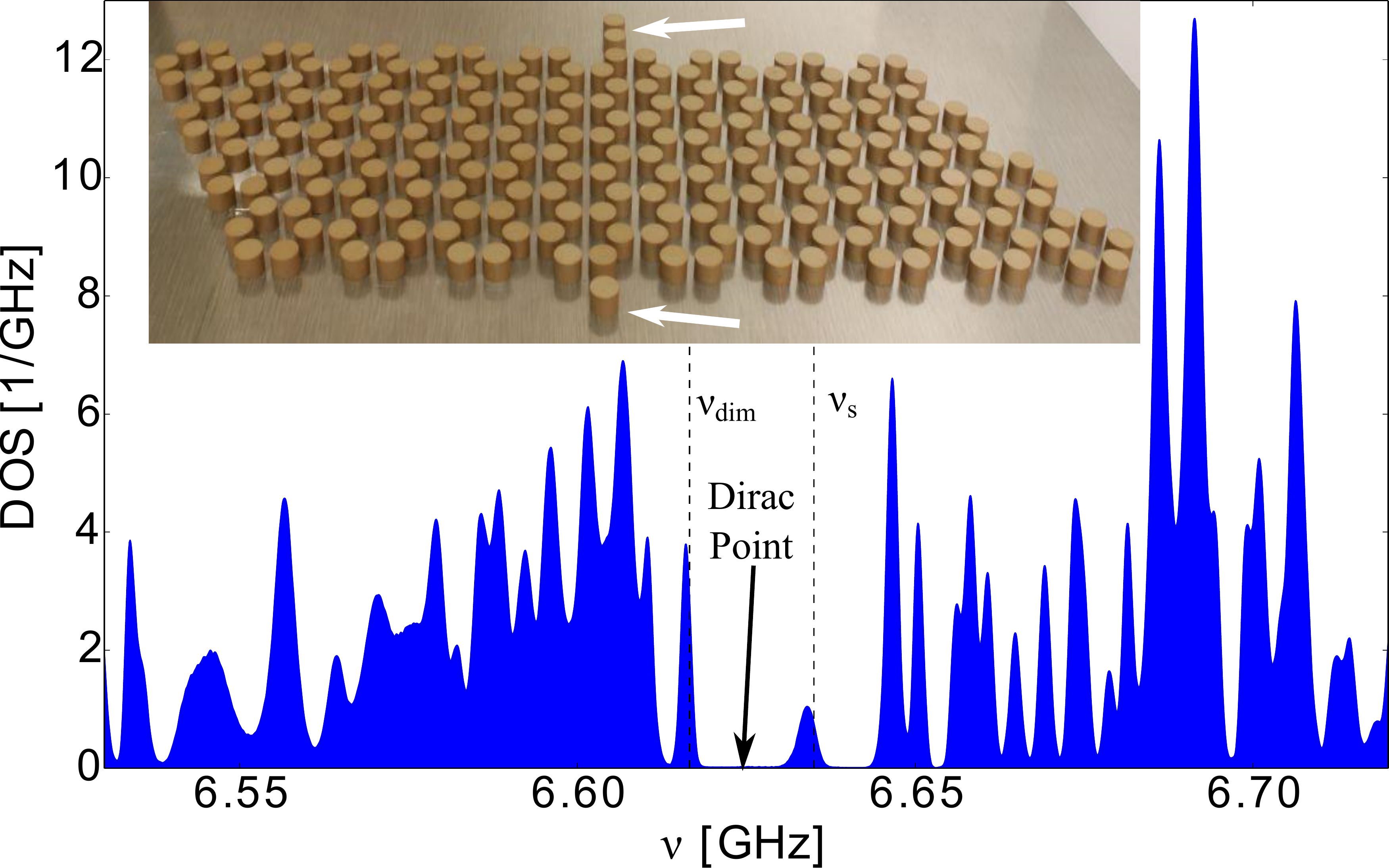}\\
\caption{\label{fig:GraphenePhotoDos} (color online).
(Inset) A photograph of the artificial graphene flake, including the supporting metallic plate and the perturber resonators at the boundary (white arrows). The graph shows the DOS of the unperturbed flake (for details, see Ref.~\cite{bel13b}), i.e., without a single resonator and dimer attached. The resonance frequencies of the single resonator $\nu_\mathrm{sr}$ and the dimer $\nu_\mathrm{dim}$ are marked by the dashed vertical lines.}
\end{figure}

\begin{figure}
  \includegraphics[width=\linewidth]{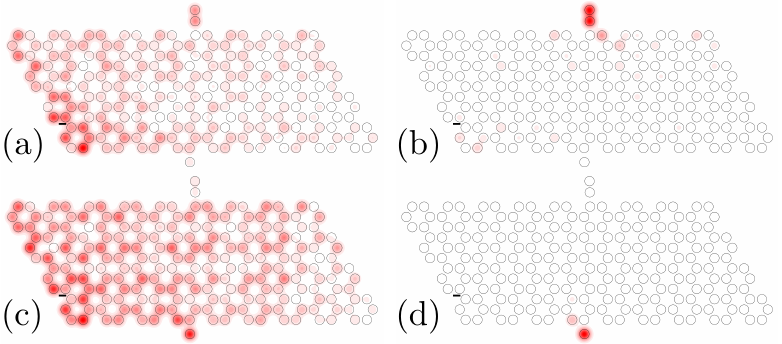}\\
\caption{\label{fig:GrapheneMonomerDimerSearch} (color online).
(a) and (c): The initial lattice state at $t=0$ for two different initial frequency ranges is shown (a) for the range around the lower lattice state $\nu_{l}$ including the dimer frequency $\nu_\mathrm{dim}$ and (c) the range around $\nu_s$ including  $\nu_{u}$. (b) and (d): The illuminated
perturber state at the search time $t=T_\mathrm{dim}$ and $T_{s}$, respectively. The resonators are indicated by the black circles and the color code corresponds to the intensity $P(r,t)$ (dark red: high probability and white: low probability).
The color code is rescaled to the maximal value.
See the Supplemental Material \cite{SupMat} for a video showing the full dynamics.}
\end{figure}


\emph{Experiments on artificial graphene.}---The experimental microwave setup is shown in the inset of Fig.~\ref{fig:GraphenePhotoDos}.
A metallic plate supports ceramic cylinders of height $h=5$\,mm and radius $r=4$\,mm which have a high index of refraction ($n \approx 6$), thus acting as resonators for a transverse electric resonance, called TE$_1$, at $\nu_0 \approx 6.65$\,GHz.
The system is closed from above by a metallic top plate at a distance of $h_p=16$\,mm (not shown).
We form a ``graphene flake'' by positioning 216 resonators in a hexagonal lattice (see inset of Fig.~\ref{fig:GraphenePhotoDos}), thus creating artificial graphene \cite{kuh10a,bel13a,bel13b}.
The top plate holds a loop antenna coupling via the magnetic field into the TE$_1$ mode which can be positioned arbitrarily in the $xy$ plane above each resonator. A kink antenna is placed at a fixed position.
For details on the experimental setup and its relations to a tight-binding model, refer to Ref.~\cite{bel13b}.
In particular, the reflection of the movable loop antenna at the center of the resonator is proportional to the intensity of the eigenmodes $|\Psi(r_n)|^2$, where $n$ labels the resonator.
These reflection measurements determine the local density of states (LDOS), and by integrating over $n$, the DOS as well (for details refer to \cite{bel13b}).
A transmission measurement from the fixed antenna 1 positioned at $r_0$ to the movable antenna 2 positioned at $r_n$ yields amplitudes as well as phases.
We conduct the experiment in the frequency domain, but the results can be converted from the frequency to the time domain without any loss of information.
The DOS for the artificial graphene flake close to the Dirac point is shown in Fig.~\ref{fig:GraphenePhotoDos}.
One can clearly identify two isolated states, which are extended lattice states.
The lower one is denoted by $\nu_l$ and the upper one by $\nu_u$.
We will use the frequencies of these states near the Dirac point as working points for our search algorithm.
The boundary has been chosen to contain only armchair edges, which do not create any edge states \cite{bel14}. Edge states are typically present at the Dirac frequency and could potentially disturb the search effect.
In the next step, we attach two perturbations at the side of the flake; see the inset of Fig.~\ref{fig:GraphenePhotoDos}.
The perturbation in the foreground is a single resonator with eigenfrequency $\nu_{s}$ close to $\nu_u$, the one in the background consists of
a dimer, that is, two strongly coupled resonators, with its lower frequency $\nu_\mathrm{dim}$ adjusted to $\nu_l$ (see also the dashed lines in Fig.~\ref
{fig:GraphenePhotoDos}). We have chosen this dimer configuration for having a parameter to fine-tune the frequency $\nu_\mathrm{dim}$.
The perturbations induce new resonance states interacting with the lattice states $\nu_l$ and $\nu_u$.
This setup allows for a search at two {\em different} frequencies---the search thus acts as a sensitive switch in frequency space.

We now obtain the transmission function $S_{12}(r_0,r_n,\nu)$ measured at every position $r_n$ and
frequency $\nu$; the time dynamics yielding the search is obtained by Fourier transformation (FT) of $S_{12}(r_0,r_n,\nu)$ in a small frequency
window around either $\nu_l$ or $\nu_u$. Such a FT corresponds to a time dynamics induced by a pulse of the form $\exp(i \nu_it) \sin
(\Delta_i t)/t$ at the position $r_0$ of antenna 1, where $i = l$ or $u$ and $\Delta_i$ corresponds to the frequency window around
$\nu_i$. Note that the pulse at a given $\nu_i$ is independent of the position of the resonator to be searched for and will find this resonator wherever it is positioned.
We would like to emphasize that the signal obtained by FT is completely {\em equivalent} to a direct measurement in the time domain using a microwave pulse generator \cite{ste95,arXpol14}.
Figures~\ref{fig:GrapheneMonomerDimerSearch}(a) and \ref{fig:GrapheneMonomerDimerSearch}(c) show the thus obtained initial state $P(r_n,t)=|FT[S_{12}(r_n,\nu)]|^2$ at $t=0$ for the two different initial frequency windows.
The state shown in Fig.~\ref{fig:GrapheneMonomerDimerSearch}(a) involves only frequencies close to $\nu_l$; in Fig.~\ref{fig:GrapheneMonomerDimerSearch}(c) only frequencies close to $\nu_u$ are included.
After some time $t=T_\mathrm{dim}$ ($t=T_{s}$), the dimer (the single resonator) is illuminated and the perturbation is thus found [see Figs.~\ref{fig:GrapheneMonomerDimerSearch}(b) and \ref{fig:GrapheneMonomerDimerSearch}(d)].
The search times $T_\mathrm{dim}$ and $T_{s}$ are slightly different due to differences in the coupling of the perturbation to the corresponding extended states.

By working near the Dirac point, the search can be extended to an arbitrary number of resonators $N$ in principle;
in praxis, the size of our model system is limited due to the overall absorption (a quality factor around 1000).
A detailed analysis of how the search time scales with $N$ is thus not possible here.
We will demonstrate this in a slightly different setup using a linear chain at the end of the Letter.

Our experiment demonstrates that the effect of spatial searching can be achieved in a graphenelike setup with tight-binding interaction.
This is encouraging as the model here  differs significantly from the theoretical study presented in Ref.~\cite{FGT14} due to different boundary
conditions, experimental uncertainties and absorption. Our results point towards a completely new set of applications in the actual carbon
material---graphene---where the limitations due to absorption are less severe. In addition to forming the basis of a device with fast
searching facility, as demonstrated above, the results presented in Fig.~\ref{fig:GrapheneMonomerDimerSearch} can also be interpreted in the framework of directed electron transport.
The device can be used as a sensitive switch, where current will be directed either to the upper (dimer) or lower (single resonator) port by a small shift in the carrier frequency of the input pulse.

\begin{figure}
  \includegraphics[width=\linewidth]{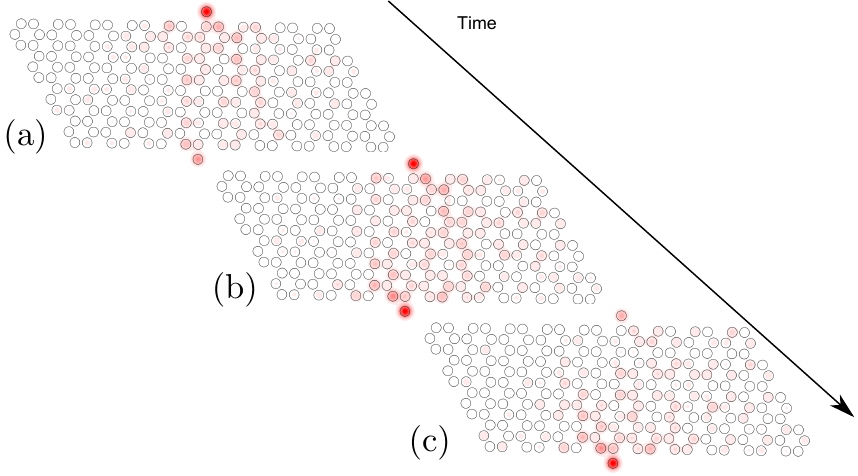}\\
\caption{\label{fig:GrapheneQuantumState} (color online).
Intensity propagation in an artificial graphene flake with two equal perturbing resonators attached. The initial state at $t=0$ is localized at the upper resonator and is transferred via a lattice state to the state sitting on the opposite resonator. The presentation is as in Fig.~\ref{fig:GrapheneMonomerDimerSearch}.
See the Supplemental Material \cite{SupMat} for a video showing the full dynamics.}
\end{figure}

Spatial quantum search can also be used for communication and ``quantum state transfer'' \cite{Nik14,HT09,FGT14} by perturbing the lattice with two equivalent resonators whose eigenfrequencies are both tuned to a single eigenfrequency of the unperturbed lattice. This scenario can also be interpreted as a search starting from one resonator and finding the other similar to the search algorithm presented in Refs.~cite{rom07,rom09}. The perturbers are attached at two different positions to the lattice and interact only via the lattice state.
Launching an initial pulse at $r_0$ will illuminate both resonators equally, but with reduced brightness.
If we, however, prepare the initial state in one of the perturbing resonators and let the system evolve thereafter, the pulse will actually
travel from the initial resonator via the lattice to the second resonator. The evolution is presented in Fig.~\ref{fig:GrapheneQuantumState}.
This time the upper resonators [Fig.~\ref{fig:GrapheneQuantumState}(a)] is illuminated initially; we then go to a state living both in the lattice and
on the two perturbing resonators [Fig.~\ref{fig:GrapheneQuantumState}(b)].
Then the other resonator lights up [Fig.~\ref{fig:GrapheneQuantumState}(c)] and (nearly) the entire amplitude is transferred from one resonator to
the other. This opens the way towards directed signal transfer and control in graphene. However, graphene cannot yet be manipulated on a
single atom level, as would be necessary for making use of the effects described in this Letter; our results may guide future
research efforts in this direction.


\begin{figure}
\includegraphics[width=0.75 \linewidth]{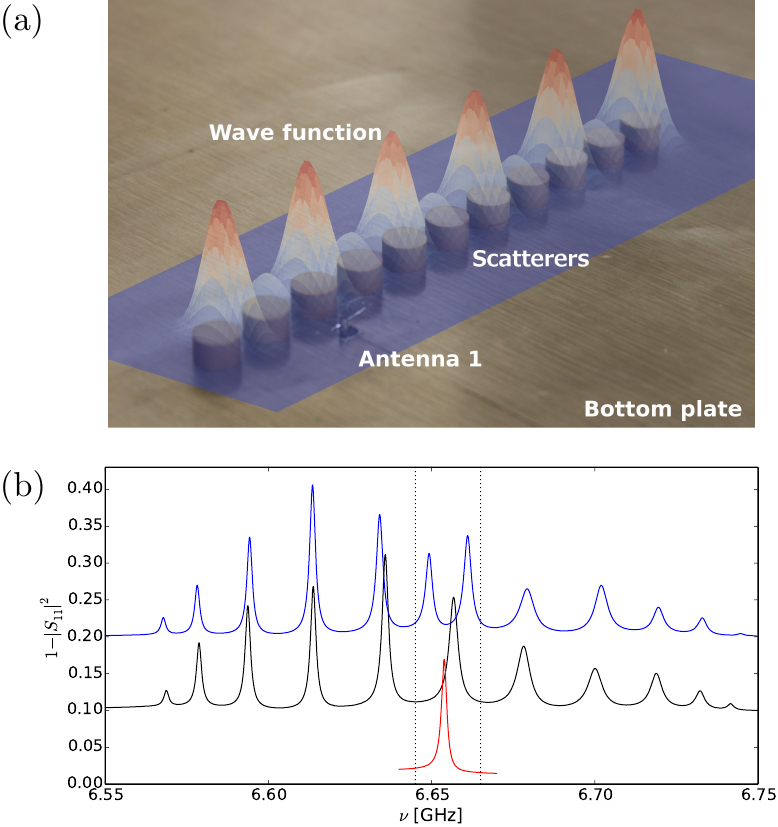}
  \caption{ \label{fig:ExpSetupWv} (color online).
  (a) Experimental setup of the linear chain with $N$=11 (without a perturber). The top plate and antenna are not shown. The wave function corresponding to the central resonance is superimposed.
  (b) Reflection measurement of a single resonator (red line, bottom), a regular chain with 11 resonators (black line, center), and a single resonator attached to the chain (blue line, top). The dotted lines mark the frequency range used for the Fourier transform (baselines shifted).}
\end{figure}

\emph{Experiment on linear chains.}---We demonstrate the $\sqrt{N}$ scaling behavior of the search time in our experimental setup on a quasi-one-dimensional system,
a linear chain. Searching is possible here only for small $N$'s, as the distance between neighboring resonances scales like $N^{-2}$ and the
number of resonances will flood the avoided crossing eventually for $N > N_\mathrm{cut}$ \cite{FGT14}.
A photograph of the unperturbed chain with $N=11$ resonators is depicted in Fig.~\ref{fig:ExpSetupWv}(a),
that is, well below the cut-off $N_\mathrm{cut}=27$ for the setup shown here. The reflection spectrum $1-|S_{11}|^2$ of the unperturbed chain containing 11 resonances is shown in Fig.~\ref{fig:ExpSetupWv}(b) (black line, central spectrum). The central resonance frequency for a chain with an odd number of resonators is always at the eigenfrequency of the single resonator. The lattice mode corresponding to the central frequency is used for the wave search; it is superimposed on the photograph.
When attaching a perturber resonator, we find an additional resonance interacting with the lattice state resulting in resonance splitting (blue line, top spectrum).
The propagator is calculated by Fourier transforming the frequency range marked by the dotted vertical lines in Fig.~\ref{fig:ExpSetupWv}(b).

\begin{figure}
  \includegraphics[width= 0.75 \linewidth]{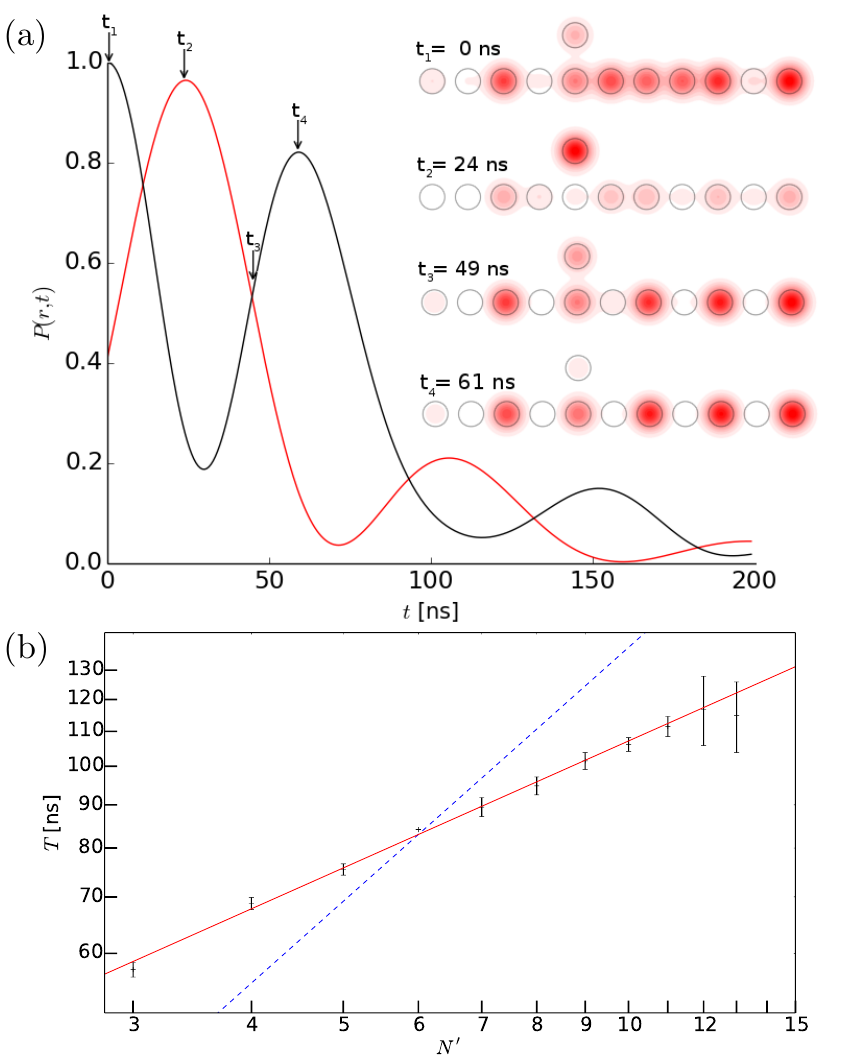}\\
  \caption{ \label{fig:GroverSearchSingleDisk} (color online).
  (a) The intensity $P(r,t)$ of the chain integrated over the 11 resonators (black line) and of the perturber (red line) is shown. The signals are normalized such that the initial state in the chain at $t$=0 is 1. The corresponding intensity distributions for specific times are shown as insets.
  See the Supplemental Material \cite{SupMat} for a video showing the full dynamics.
  (b) The beating time as a function of $N^\prime$; solid (red) line: square root dependance; dashed (blue) line: linear increase. The error bars indicate the time difference between different recurrences.}
\end{figure}

Figure~\ref{fig:GroverSearchSingleDisk}(a) depicts the temporal $P(r,t)$ behavior of the intensity of the whole chain (black line) and of the perturber
resonator (red line). The initial state ($t=0$) is adjusted to the maximal amplitude on the lattice and normalized to 1.
We observe an oscillatory behavior corresponding to a beating between the two states, where the search time is given by half of the beating time
$T$. The insets show the corresponding intensities of the wave function at the indicated times.
Figure~\ref{fig:GroverSearchSingleDisk}(b) shows the time $T$ as a function of sites $N^\prime=(N+1)/2$ for a different number of odd sites $N$ ranging from 5 to 27.
We observe the predicted $\sqrt{N^\prime}$ behavior.


\emph{Conclusion.}---We have demonstrated a first proof-of-principle experiment of a continuous quantum wave search on an (artificial) graphene lattice.
The search is facilitated by bringing a lattice state into resonance with a localized perturber state at an avoided crossing.
Apart from searching perturber states, one can also use this scheme to address particular sites using a frequency scan, thus initiate a switching behavior, and to transfer signals between sites without knowing their positions.
The experiment is limited mainly by losses and absorption. Reducing the resonance widths further, i.e., obtaining a better quality factor $Q$, can be realized using coupled supraconducting cavities (quality factors of about $Q\approx 10^8$ are possible).

%
We would like to acknowledge our inspiring discussion with Klaus Richter and Olivier Legrand.

\end{document}